\documentclass[pra,showkeys,showpacs]{revtex4}
\usepackage{amsthm}
\usepackage{amsmath}
\usepackage{amssymb}
\usepackage{dsfont}
\usepackage{graphicx}

\makeatletter

\providecommand{\tabularnewline}{\\}

\newtheorem{thm}{\protect\theoremname}

\makeatother

\providecommand{\theoremname}{Theorem}

\begin{document}
\global\long\def\ket#1{\left|#1\right\rangle }
\global\long\def\bra#1{\left\langle #1\right|}
\global\long\def\braket#1#2{\left\langle #1|#2\right\rangle }
\global\long\def\mel#1#2#3{\left\langle #1|#2|#3\right\rangle }

\global\long\def\Id{\mathds{1}}
\global\long\def\R{\mathds{R}}
\global\long\def\C{\mathds{C}}
\global\long\def\T{\Theta}

\title{Sure-Success Quantum Algorithms on Weight Decision Problem}

\author{K. Uyan\i{}k}
\email{ukivanc@newton.physics.metu.edu.tr}

\author{S. Turgut}
\email{sturgut@metu.edu.tr}

\affiliation{Department of Physics, Middle East Technical University\\
 06800, Ankara, TURKEY}

\begin{abstract}
Conditions on sure-success decidability of weights of Boolean functions are
presented for a given number of generalized Grover iterations. It is shown
that the decidability problem reduces to a system of algebraic equations of
a single variable. For problems that require a large number of iterations,
it is observed that the iteration number of sure-success quantum algorithms
scale as the square root of the iteration number of the corresponding
classical probabilistic algorithms. It is also demonstrated that for a few
iterations, quantum algorithms can be more efficient than this.
\end{abstract}



\maketitle

\section{Introduction}

\label{sec:Introduction}

Quantum algorithms have been proved to be exponentially faster than their
classical counterparts in problems such as Deutsch's
problem\cite{Deutsch85,Deutsch92}, Simon's problem\cite{Simon94} and
super-polynomially faster in order finding\cite{Shor94}. All of these
problems have some simplifying features and the quantum algorithms which
solve these problems make use of such features while exploiting quantum
parallelism and, in some cases, entanglement. If there is no such simplifying
feature, quantum algorithms do not perform this well. Nevertheless, as Grover
showed, one can obtain at least a quadratic speedup\cite{Grover97,Grover98}
for searching a single item in an unstructured database. Unfortunately this
small-scale speedup is shown to be the upper limit of quantum database search
algorithms\cite{Bennett97,Boyer98,Zalka99}; however wide range of
applications compensate for this.

Several generalizations and variations of Grover's algorithm is explored up
to now. In Ref.~\onlinecite{Boyer98}, the problem of searching for several
items, instead of a single item, is studied. In this case, the algorithm
gives one of the solutions randomly at the output. In
Ref.~\onlinecite{Grover98}, it is shown that using an arbitrary unitary
(instead of the Hadamard transform which were used in the original Grover
algorithm) does not change the $O(\sqrt{N})$ run-time as long as it is used
consistently. If there is an inner structure to be exploited, this unitary
can be chosen accordingly to obtain quadratic speedup for the new search
space. Biham \textit{et al.} further generalized the algorithm such that the
initial state\cite{Biham99} and the phase inversion angle\cite{Biham00} are
arbitrary and obtained similar results. Grover's original algorithm is not
deterministic, but with a few tweaks it is possible to obtain solutions with
zero error probability\cite{Hoyer00,Long01}. It does not monotonically
converge to a solution (i.e., if you run it too much, it misses the target)
but it can also be altered to become a fixed point algorithm\cite{Tulsi06}.

An important problem related to database search is counting the number of
roots of a given Boolean function $f$. In this problem, the aim is to find
the number of inputs $x$ that gives $f(x)=0$ (or equivalently the number of
the roots of $f(x)=1$ is sought). Brassard \textit{et al.} gave a practical
method to accomplish this task approximately, by employing Grover iteration
as well as quantum Fourier transform\cite{Brassard98}. Another specific
problem in which Grover search iteration is used is the weight analysis of
Boolean functions. The ratio of the number of solutions of $f(x)=1$, to the
number of all possible inputs is called the weight of a function. Weight
analysis of Boolean functions has proved to be useful in areas such as
cryptanalysis\cite{Filiol98}, coding theory\cite{MacWilliams96},
fault-tolerant circuit design\cite{Chakrabarty96}, and for built-in
self-testing circuits\cite{Chakrabarty95}. Weight analysis using Grover
algorithm is studied by Braunstein \textit{et al.} and Choi and Braunstein in
a series of papers\cite{Braunstein07,Choi11}. They first solve the problem
with a restriction where they only consider two ``symmetric'' weights (i.e.,
the function is known to have a weight either equal to $\rho_{1}$ or
$\rho_{2}=1-\rho_{1}$). Then they generalize it to the asymmetric case where
the restriction on the weights is now removed. This task can also be
accomplished by using quantum counting\cite{Brassard98}. However, in that
case, one needs to introduce quantum Fourier transform and quantum counting
is still slightly slower than the weight decision algorithm.

In this contribution, the weight decision problem is studied by an
alternative approach. The main motivation is to see which weights could be
distinguished by a given number of function evaluations, especially in the
regime where only a few evaluations are required. As the algorithm devised by
Braunstein \emph{et al.} requires at least 3 function evaluations, the
approach used in this article covers an unexplored territory. The
organization of the article is as follows. Section \ref{sec:Definitions}
starts with a general Grover iteration which consists of unitaries which are
more general than the Hadamard transform of the original algorithm. The basic
definitions are given and an essential theorem is presented in this section.
The first problem tackled in Section \ref{sec:Main-Results} is the decision
problem of a zero weight and a non-zero weight. After that, the decision
problem of two non-zero weights by a few iterations is discussed. Exact
solutions are given for 1 and 2 iterations. For 3 or more iterations, the
associated equations become too complicated and thus we are forced to present
numerical solutions for a few cases. A brief comparison of classical and
quantum efficiencies and a comparison to the algorithms of Braunstein
\emph{et al.} are also included in this section. Finally, brief concluding
remarks are given in section \ref{sec:Conclusions}.

\section{Preliminaries and the Construction of the Problem}
\label{sec:Definitions}

Let us first define the problem. We are given a Boolean function $f$ of $N$
possible inputs $\left(f:\{0,1,\cdots,N-1\}\longrightarrow\{0,1\}\right)$. We
somehow know that the function has either the weight $\rho=r/N$ or the weight
$\rho^{\prime}=r^{\prime}/N$. In other words, the total number of inputs $x$
for which $f(x)=1$ is either $r$ or $r^{\prime}$. Our job is to determine
which one is the case. Note that by ``weight decision problem'' we imply the
general case which is usually ``asymmetric'' as in \cite{Choi11}, that is to
say, $r+r^\prime$ is not necessarily equal to $N$.

The evaluation is implemented in a quantum computer as a black box. The
function evaluator, upon reading the input register (which is an $N$-level
system), adds the value of the function on the result qubit. Denoting the
unitary transformation of the function evaluation by $U_{f}$ we have
\begin{equation}
U_{f}\ket{x}_I \otimes \ket{b}_{R} = \ket{x}_I \otimes \ket{b\oplus f(x)}_{R} \enskip,
\label{eq:OracleDefinition}
\end{equation}
where I denotes the $N$-level input register and R denotes the result qubit.
Using the basis $\ket{\pm}_{R}=\left(\ket 0_{R}\pm\ket
1_{R}\right)/\sqrt{2}$, the action of $U_f$ can also be expressed as
\begin{equation}
U_{f}\ket{x\pm}_{IR}=\left(\pm 1\right)^{f(x)} \ket{x\pm}_{IR}\enskip.
\end{equation}
Therefore, when $f(x)=1$, the overall phase of the state $\ket{x-}$ is
rotated by $\pi$ radians. The phases of the states $\ket{x+}$ remain
unchanged irrespective of the function $f$.

Let A be an ancilla system. Let $\ket{\beta_1},\ldots,\ket{\beta_n}$ be a set
of orthonormal vectors in the state space of the composite system AIR.
Consider the following unitary operator $S$ that acts on the composite system
\begin{equation}
S\equiv\Id-2\sum_{i=1}^{n}\ket{\beta_{i}}\bra{\beta_{i}}\label{eq:Smatrix}\enskip.
\end{equation}
This operator is essentially an inversion operation in an $n$-dimensional
subspace of the Hilbert space, namely the subspace spanned by
$\{\ket{\beta_i}\}_i$. The successive application of $U_{f}$ and $S$
constitute a single iteration step $Q_{f}\equiv-S(\mathds{1}_A\otimes
U_{f})$. To express the effect of multiple iterations of $Q_f$ on an
arbitrary initial state, it is convenient to first define an  $n\times n$
matrix $\mathbb{C}$, which will be called as the ``cosine matrix'', as
\begin{equation}
\C_{ij}\equiv\mel{\beta_{i}}{\mathds{1}_A\otimes U_{f}}{\beta_{j}}\enskip.\label{eq:Cmatrix}
\end{equation}
As $U_f$ is unitary with real eigenvalues of $\pm1$, it is also hermitian.
This implies that the cosine matrix $\C$ is hermitian and all of its
eigenvalues are in the $[-1,+1]$ interval. In that case, we can think of $\C$
as the cosine of an angle matrix $\T$, i.e., $\C=\cos\T$. Next, we define
$n\times n$ matrices $\R^{(m)}$  for all integer values $m$ by
\begin{equation}
\R^{(m)}\equiv\frac{\sin(m\T)}{\sin(\T)}\enskip,
\label{eq:Rmatrix}
\end{equation}
which is actually a polynomial function of $\C$. Note that all of these
matrices depend on the function $f$. When necessary, this dependence will be
shown by $\C(f)$, $\T(f)$ and $\R^{(m)}(f)$. But, the function will not be
shown explicitly when there can be no confusion.

We can express our fundamental result in terms of the $\R^{(m)}$ matrices as
follows.
\begin{thm}
\label{thm:Qp} If the initial state is one of $\ket{\beta_i}$, then, after
$m$ iterations of $Q_f=-S(\mathds{1}_A\otimes U_f)$, the final state is
\begin{equation}
Q_f^{m}\ket{\beta_{i}}=\sum_{j=1}^n\left(\ket{\beta_{j}}\R_{ji}^{(m+1)}-U_{f}\ket{\beta_{j}}\R_{ji}^{(m)}\right)\enskip.
\label{eq:ThmQp}
\end{equation}
\end{thm}

The proof is rather straightforward. One first verifies Eq.~\eqref{eq:ThmQp}
for $m=0$ and $m=1$. After that, showing that the matrices $\R^{(m)}$ satisfy
the following recurrence relation
\begin{equation}
 \R^{(m+2)}-2\C\R^{(m+1)}-\R^{(m)}=0\enskip,
\end{equation}
and using these in $Q_f^{m+1}\ket{\beta_i}$ completes the proof of the
theorem. The details are left to the reader. A slightly more complicated, but
still simple expression can be found for $Q_f^m\ket{\psi}$ for any arbitrary
initial state $\ket{\psi}_{AIR}$. However, that general case will not be
needed in this contribution.

The problem is as follows. We prepare the initial state of the input register
I, result qubit R and ancilla A in the state
\begin{equation}
  \ket{\beta}\equiv \sum_{i=1}^n c_{i}\ket{\beta_{i}}
\end{equation}
where the amplitudes $c_i$ will be determined later. After that, the
evaluation of the unknown function $f$ and the unitary $S$ are alternatingly
applied $m$ times. In other words, $Q_f^m$ is applied on the composite
system. The final state of the composite system AIR is
\begin{equation}
  \ket{\Phi_{f}} \equiv Q_{f}^{m}\ket{\beta}\enskip.
\end{equation}
Finally, a measurement is carried out on the composite system AIR for
determining the weight of the function $f$.

If this final measurement enables us to measure the correct weight of the
function $f$ deterministically (with probability 1), then, it is necessary
that all final states corresponding to functions with different weights are
orthogonal. In other words, if $f$ and $g$ are any two functions that could
be possibly computed by the black box device, we should have
$\braket{\Phi_{f}}{\Phi_{g}}=0$ whenever $f$ and $g$ have different weights.
In this contribution, only the case where the function computed by the black
box device has either the weight $\rho$ or $\rho^\prime$ will be considered.
Therefore, if the correct weight of the unknown function could be determined
after $m$ iterations, then the set of final states $\ket{\Phi_f}$ for
functions with weight $\rho$ and the corresponding set for functions with
weight $\rho^\prime$ should be in orthogonal subspaces.

Undoubtedly, how the final measurement is carried out is also important from
the computation point of view. However, in this contribution, the primary
concern is the possibility of distinguishing functions with different weights
and not how the steps of the algorithm can be implemented. Moreover, the
length of the algorithm will be measured with the number of evaluations of
the unknown function $f$, presumably because this is very costly. Hence, it
is assumed that the preparation of the initial state $\ket{\beta}$,
implementation of the unitary $S$ and the final measurement requires a much
smaller number of computation steps than carrying out $U_f$. For this reason,
we are inclined to find only the smallest iteration number $m$ that is
necessary for distinguishing two given weights $\rho$ and $\rho^\prime$.

For the weight decision problem, it appears that the following choice for the
$\ket{\beta_i}$ is sufficient.
\begin{equation}
\ket{\beta_{i}}=\ket{\beta_{i}}_{AIR} \equiv \ket{\alpha_{i}}_{A} \otimes
\left(\sqrt{\mu_{i}}\frac{1}{\sqrt{N}}\sum_{x=0}^{N-1}\ket{x-}_{IR}+\sqrt{1-\mu_{i}}\ket{0+}_{IR}\right)
\label{eq:beta_defined}
\end{equation}
Here, $\mu_{i}$ are real parameters between 0 and 1 and the ancilla states
$\ket{\alpha_{i}}_{A}$ are normalized and mutually orthogonal to each other
(i.e., $\braket{\alpha_{i}}{\alpha_{j}}=\delta_{ij}$), so that
$\ket{\beta_{i}}_{AIR}$ are also normalized and mutually orthogonal to each
other. In that case, the cosine matrix $\C$ is diagonal. If the unknown
function $f$ has weight $\rho$, then the $i$th diagonal entry of $\C$ is
\begin{equation}
  \C_{ii} = \cos\theta_i(f) =\mel{\beta_{i}}{\mathds{1}_A\otimes U_{f}}{\beta_{i}} =  1-2\rho\mu_i
\end{equation}
The angle eigenvalues are therefore bounded in the interval
\begin{equation}
  0 \leq \theta_i(f)\leq \arccos(1-2\rho)\enskip.
\end{equation}
The matrices $\R^{(m)}$ are also diagonal in this case and their $i$th
diagonal entry are given by
\begin{equation}
  \R_{ii}^{(m)} = \frac{\sin m\theta_i(f)}{\sin\theta_i(f)} \enskip.
\end{equation}

\section{Results}
\label{sec:Main-Results}

\subsection{Distinguishing zero weight functions from non-zero weight functions}
First, the case where one of the weights is identically zero, $\rho^\prime=0$
is investigated. In this case, the decidability of distinguishing a function
$f$ with weight $\rho=\dfrac{r}{N}$ and the zero function $z$ (which is
defined as $z(x)=0$ for all $x$) is studied. In other words, we are given a
function $f$ and we are told that either the function $f$ has weight $\rho$
(but otherwise arbitrary) or it is the zero function. We are asked to
determine if $f$ is the zero function or not with a minimum possible number
of function evaluations. Classically, the deterministic algorithms require
$N(1-\rho)+1$ evaluations in the worst case although a single evaluation is
sufficient if we are lucky. It is of some interest to see how quantum
algorithms perform for this problem.

In this case, $U_{z}=\Id$ and $Q_{z}\ket{\beta_{i}}=\ket{\beta_{i}}$, thus
leading to $Q_{z}^{m}\ket{\beta_{i}}=\ket{\beta_{i}}$. If all functions with
weight $\rho$ could be distinguished from $z$ after $m$ function evaluations,
then we should have $\braket{\Phi_{z}}{\Phi_{f}}=0$ for all functions $f$
with weight $\rho$.
\begin{align}
0 &=\braket{\Phi_{z}}{\Phi_{f}}  =  \mel{\beta}{Q_{f}^{m}}{\beta} \\
 & =  \sum_{ij} a_{i}^* \bra{\beta_{i}} Q_{f}^{m}\ket{\beta_{j}}a_{j} \label{eq:InnerProductPth}\\
 & =  \sum_{ij}a_{i}^*\left(\R^{(m+1)}(f)-\C(f)\R^{(m)}(f)\right)_{ij}a_{j}\\
 & =  \sum_{ij}a_{i}^*\left(\cos m\T(f) \right)_{ij}a_{j} \enskip. \label{eq:ZeroFunction}
\end{align}
For any possible iteration number $m$, the equation above can be satisfied
with $n=1$ (i.e., $S$ is an inversion in a one-dimensional subspace spanned
by $\ket{\beta_1}$). In this case, $\Theta(f)$ is a $1\times1$ matrix, which
we may denote by the value $\theta_1(f)$. We should therefore have $\cos
m\theta_1(f)=0$ for all functions $f$ with weight $\rho$. If the iteration
number $m$ is the minimum possible value, then $\theta_1(f)$ should be
independent of $f$ (the definition of $\ket{\beta_1}$ in the form in
Eq.~\eqref{eq:beta_defined} is consistent with this) and should be given by
$\theta_1(f)=\pi/2m$. The minimum iteration number is, therefore, the
smallest integer $m$ where we can find a number $\mu_1$ in $[0,1]$ interval
such that
\begin{equation}
  \cos\frac{\pi}{2m}=1-2\rho\mu_1\enskip.
\end{equation}
It is then straightforward to show that the smallest iteration number is
given by
\begin{equation}
  m_{min}(\rho) = \left\lceil \frac{\pi}{2\arccos(1-2\rho)} \right\rceil\enskip,
\end{equation}
where $\lceil y\rceil$ denotes the smallest integer greater than or equal to
$y$.
\begin{table}[h]
\begin{tabular}{ccc}
\hline
$m$ &  & $\rho_{\textrm{min}}(m)$\tabularnewline
\hline
\hline
$1$ &  & $0.5$%
\footnote{A special variation of Deutsch-Jozsa algorithm%
}\tabularnewline
$2$ &  & $0.15$\tabularnewline
$3$ &  & $0.067$\tabularnewline
$4$ &  & $0.038$\tabularnewline
$5$ &  & $0.024$\tabularnewline
$10$ &  & $0.0062$\tabularnewline
\hline
\end{tabular}
\caption{Minimum weights of functions, which can be distinguished from the zero function $z$ by the only $m$ function evaluations.}
\label{tab:MinWeightDistFromZeroF}
\end{table}

One can also ask the reverse question: which weights $\rho$ can be
distinguished by $m$ iterations? In that case, the condition on the weights
is found to be
\begin{equation}
\rho \geq \rho_{\textrm{min}}(m)=\frac{1}{2}\left(1-\cos\frac{\pi}{2m}\right)\enskip.
\end{equation}
For a few small $m$ values, the threshold weights $\rho_{\textrm{min}}$ are
tabulated in Table~\ref{tab:MinWeightDistFromZeroF}.

It appears that, with a single function evaluation (i.e., $m=1$) any function
$f$ with weight $\rho$ can be distinguished from the zero function $z$
provided that $\rho\geq1/2$. This special case corresponds to a variation of
the Deutsch-Jozsa problem\cite{Deutsch92}. In the Deutsch-Jozsa problem, one
needs to distinguish functions $f$ with weight $\rho=1/2$, from the constant
functions $z$ and $u$ where $u=z\oplus1$ (i.e., $u(x)=1$ for all $x$).

If the weight $\rho$ is smaller than $1/2$, more than one function
evaluations are necessary. As it will be discussed below, when the weights to
be distinguished $\rho$ and $\rho^\prime$ are closer to each other, more
function evaluations are needed to identify the weight correctly. This is
also the case for the current problem: when $\rho\ll1$, one needs
$m\sim\pi/4\sqrt{\rho}$ function evaluations in order to distinguish the
weight $\rho$ from the weight $\rho^\prime=0$.

\subsection{Distinguishing two non-zero weights}

Now, consider the problem of identifying the weight of the function when both
of the possible weights $\rho$ and $\rho^\prime$ are non-zero. For
simplicity, let us consider the choices in Eq.~\eqref{eq:beta_defined} so
that the angle matrices $\T$ are diagonal. Let $f$ and $g$ be two functions
with respective weights $\rho$ and $\rho^\prime$. The inner product of the
final states is given by
\begin{eqnarray}
\braket{\Phi_{f}}{\Phi_{g}} & = & \sum_{i=1}^{n}\left\vert c_{i}\right\vert^{2}
       \bigg(  \R_{ii}^{(m+1)}(f)\R_{ii}^{(m+1)}(g)-\R_{ii}^{(m+1)}(f)\R_{ii}^{(m)}(g)\cos\theta_{i}(g)\nonumber \\
    &  & -\R_{ii}^{(m)}(f)\R_{ii}^{(m+1)}(g)\cos\theta_{i}(f)+\R_{ii}^{(m)}(f)\R_{ii}^{(m)}(g)\cos\theta_i(f\oplus g)
           \bigg)=0\enskip.
 \label{eq:FfFgInnerProduct}
\end{eqnarray}
Consider the last term inside the sum of Eq. \eqref{eq:FfFgInnerProduct}. The
matrix element can be evaluated as
\begin{equation}
\cos\theta_i(f\oplus g)=\mel{\beta_{i}}{\mathds{1}_A\otimes U_{f\oplus g}}{\beta_{i}}=1-\mu_{i}\frac{2t}{N}\enskip,
\end{equation}
where $t$ is the number of inputs which make $f\oplus g$ one, i.e., $t/N$ is
the weight of the function $f\oplus g$. For all possibilities for $f$ and
$g$, $t$ can take on the values
$t=r-r^{\prime},r-r^{\prime}+2,\ldots,r+r^{\prime}-2,r+r^{\prime}$. As
Eq.~\eqref{eq:FfFgInnerProduct} is linear in $t$, the condition in
Eq.~\eqref{eq:FfFgInnerProduct} is reducible to two independent equations
\begin{eqnarray}
\sum_{i=1}^{n}\left|c_{i}\right|^{2}A_{i} & = & 0\nonumber \\
\sum_{i=1}^{n}\left|c_{i}\right|^{2}B_{i} & = & 0\label{eq:Set2BeSolved}
\end{eqnarray}
subject to the condition
\begin{equation}
\sum_{i=1}^{n}\left|c_{i}\right|^{2}=1\enskip,
\end{equation}
where $A_{i}$ and $B_{i}$ are given as follows:
\begin{eqnarray}
A_{i} & \equiv & \cos(m\theta_{if})\cos(m\theta_{ig})+\frac{\sin(m\theta_{if})\sin(m\theta_{ig})}{\sin\theta_{if}\sin\theta_{ig}}\big(1-\cos(\theta_{if})\cos(\theta_{ig})\big) \\\label{eq:ABexprA}
B_{i} & \equiv & \frac{\sin(m\theta_{if})\sin(m\theta_{ig})}{\sin\theta_{if}\sin\theta_{ig}}\big(2-\cos\theta_{if}-\cos\theta_{ig}\big)\enskip.\label{defn:AB}\label{eq:ABexprB}
\end{eqnarray}
Therefore we have to find $n$ tuplets $(A_{i},B_{i})$, which are just points
on the 2-dimensional plane such that
$\sum_{i}\left|c_{i}\right|^{2}\left(A_{i},B_{i}\right)=\left(0,0\right)$.
Geometrically, this means that the origin $\left(0,0\right)$ is in the convex
hull of the set of points $\left\{ \left(A_{i},B_{i}\right)\right\}$.

Note that, with the choice in Eq.~\eqref{eq:beta_defined}, $A_i$ and $B_i$
depend only on the fixed parameters $m$, $\rho$ and $\rho^\prime$; and the
adjustable parameter $\mu_i$. Keeping the dependence on the fixed parameters
as implicit, and showing the dependence on $\mu_i$ explicitly we can write
$A_i=A(\mu_i)$ and $B_i=B(\mu_i)$. Therefore, the set of points $(A_i,B_i)$
lie on a continuous curve $(A(\mu),B(\mu))$. The weights can be
distinguished, therefore, if the origin, $(0,0)$, is inside the convex hull
of the whole curve $(A(\mu),B(\mu))$ for $0\leq\mu\leq1$. In such a case, the
problem can be solved with $n=2$, i.e., one needs to find only two points on
the curve $(A(\mu),B(\mu))$ such that the line joining them passes from the
origin. The associated values of $\mu_1$ and $\mu_2$ enables us to find
$\ket{\beta_1}$ and $\ket{\beta_2}$.

Therefore, the distinguishability problem of two weights $\rho$ and
$\rho^\prime$ with $m$ function evaluations reduces to a problem in convex
analysis: determining whether a point lies within the convex hull of a curve.
Both coordinates of the curve $(A(\mu),B(\mu))$ are actually polynomial
functions of $\mu$,
\begin{align}
 A(\mu) &=T_m(y) T_m(y^\prime) + U_{m-1}(y) U_{m-1}(y^\prime)(1-yy^\prime) \\
 B(\mu) &=2(\rho+\rho^\prime) U_{m-1}(y) U_{m-1}(y^\prime)\mu
\end{align}
where $T_m$ and $U_{m}$ denote the Chebyshev polynomials that are defined by
$T_m(t)=\cos(m\arccos(t))$ and $U_{m-1}(t)=\frac{1}{m}\frac{d}{dt}T_m(t)$ and
$y=1-2\rho\mu$, $y^\prime=1-2\rho^\prime\mu$. However, determining whether
the origin lies in the convex hull of the curve is a complicated problem
which can only be solved numerically in most cases.

The case with $m=1$ is the simplest. In that case, $A(\mu)=1$ for all $\mu$
and therefore the origin can never be in the convex hull of the associated
curve. This implies that, it is not possible to distinguish two weights which
are both different from $0$ or $1$ by a single function evaluation.

\begin{figure}
\includegraphics{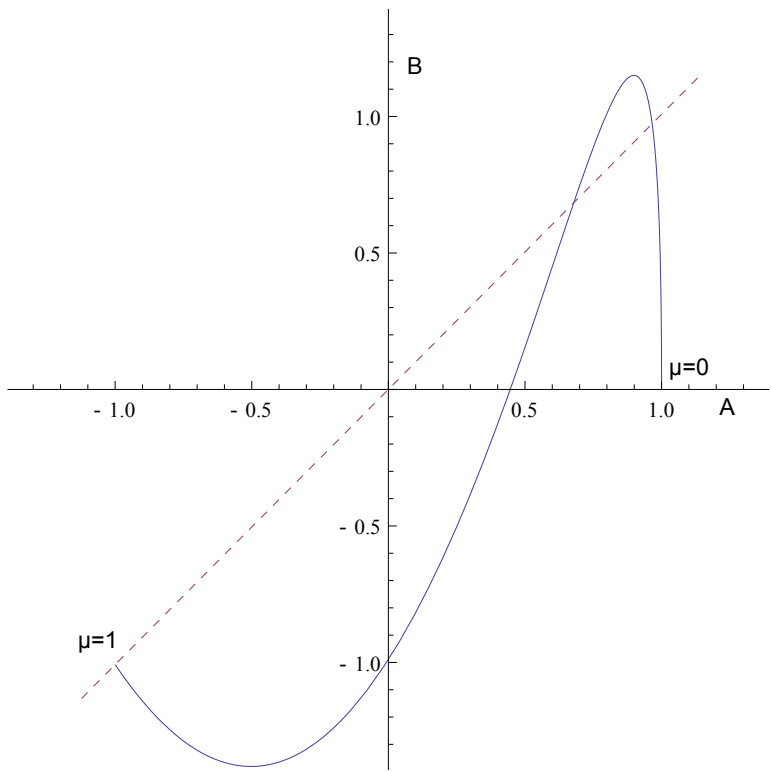}
\caption{$B(\mu)$ vs. $A(\mu)$ for $m=2$, $\rho=.95$ and $\rho^\prime=0.45$}
\label{fig:p2}
\end{figure}

The curve for the case $m=2$, $\rho=.95$ and $\rho^\prime=0.45$ is shown in
Fig.~\ref{fig:p2}. In this example, we can see that the parametric curve have
$(A,B)=(1,0)$ at $\mu=0$, which turns out to be the case for all situations.
However, where the curve ends is nontrivial. If it cuts the horizontal axis
again in the negative part of the axis, a convex combination that gives the
origin is easily achieved. This simplification covers most of the solution
space and can be utilized for a quick analysis. For a complete analysis one
has to find compact inequalities for $\rho$ and $\rho^\prime$. In order to
understand the general case, $m=2$ case is studied first.

\subsubsection{The case where $m=2$}

For $m=2,$ $A(\mu)$ and $B(\mu)$ becomes
\begin{eqnarray}
A^{(2)}(\mu) & = & 1-8\mu^{2}(\rho-\rho^\prime)^{2}\\
B^{(2)}(\mu) & = & 2\mu(\rho+\rho^\prime)(1-2\mu\rho)(1-2\mu\rho^\prime)
\end{eqnarray}
where we introduced superscripts to emphasize $m=2.$ Now we require that
there exists two points $\left(A^{(2)}(\mu_{1}),B^{(2)}(\mu_{1})\right)$ and
$\left(A^{(2)}(\mu_{2}),B^{(2)}(\mu_{2})\right)$ such that
\begin{equation}
\sum_{i=1,2}\left|c_{i}\right|^{2}\left(\begin{array}{c}
A^{(2)}(\mu_{i})\\
B^{(2)}(\mu_{i})
\end{array}\right)=\left(\begin{array}{c}
0\\
0
\end{array}\right)\enskip.
\end{equation}
To do this, we must find two values, $\mu_1$ and $\mu_2$, satisfying
\eqref{eq:p2convex00}
\begin{equation}
\frac{B^{(2)}(\mu_1)}{A^{(2)}(\mu_1)}=\frac{B^{(2)}(\mu_2)}{A^{(2)}(\mu_2)} \label{eq:p2convex00}
\end{equation}
where $B^{(2)}(\mu_1)$ and $B^{(2)}(\mu_2)$ (and, in parallel,
$A^{(2)}(\mu_1)$ and $A^{(2)}(\mu_2)$) have opposite signs. It turns out that
if there is a solution to Eq.~\eqref{eq:p2convex00}, it can be realized with
a single variable $\mu_1$, while $\mu_2$ is set to $1$. Hence, we get
\begin{equation}
  B^{(2)}(\mu)A^{(2)}(1) - A^{(2)}(\mu)B^{(2)}(1) = \left(K\mu^2 + L\mu + M\right)(\mu-1) = 0
  \label{eq:quadratic}
\end{equation}
where
\begin{eqnarray}
 K & = & 4\rho\rho^\prime\left(1-8(\rho -\rho^\prime)^2\right) \\
 L & = & 8(\rho-\rho^\prime)^2+4\rho\rho^\prime-2(\rho +\rho^\prime )\\
 M & = & (2\rho-1)(2\rho^\prime-1) \enskip.
\end{eqnarray}
Let $\Delta \equiv L^2 - 4KM$ be the discriminant of this quadratic equation.
Suppose that $\rho>\rho^\prime$. All of the following conditions have to be satisfied for being able
to distinguish these two weights:
$\rho/\rho^\prime>1+1/\sqrt{2}$, $\rho>1/2$,
$\rho -\rho^\prime > 1/2\sqrt{2}$, and $\Delta>0$.
The conditions for the case $\rho<\rho^\prime$ can be obtained simply by
using $\rho - \rho^\prime$ symmetry. These conditions are found analytically
using Eq.~\eqref{eq:quadratic} and by inspecting all possible orderings of
the zeros and poles of $B^{(2)}(\mu)/A^{(2)}(\mu)$. However, even for $m=2$,
the equation for the conditions on $\rho$ and $\rho^\prime$ requires solving
a quadratic equation. Calculations for bigger $m$ are even more forbidding
and therefore numerical computations had to be utilized instead. Numerical
results for $m=2$ case is pictured in Fig.~\ref{fig:p2p8} among further
results from $m=3$ to $m=8$.

\subsubsection{General case}

Basically the same procedure can be followed for $m \geq 2$. Solving
Eq.~\eqref{eq:p2convex00} would be more and more challenging analytically
with increasing $m$. Fortunately this problem is suitable for numerical
analysis. Note that solving this problem directly, without applying Theorem
\ref{thm:Qp} is also possible, however in that case we would have needed to
optimize $\ket{\beta_{i}}$'s. In this formulation, there is only one
parameter, $\mu$, to be optimized and this is a clear advantage.

\begin{figure}[h]
\includegraphics{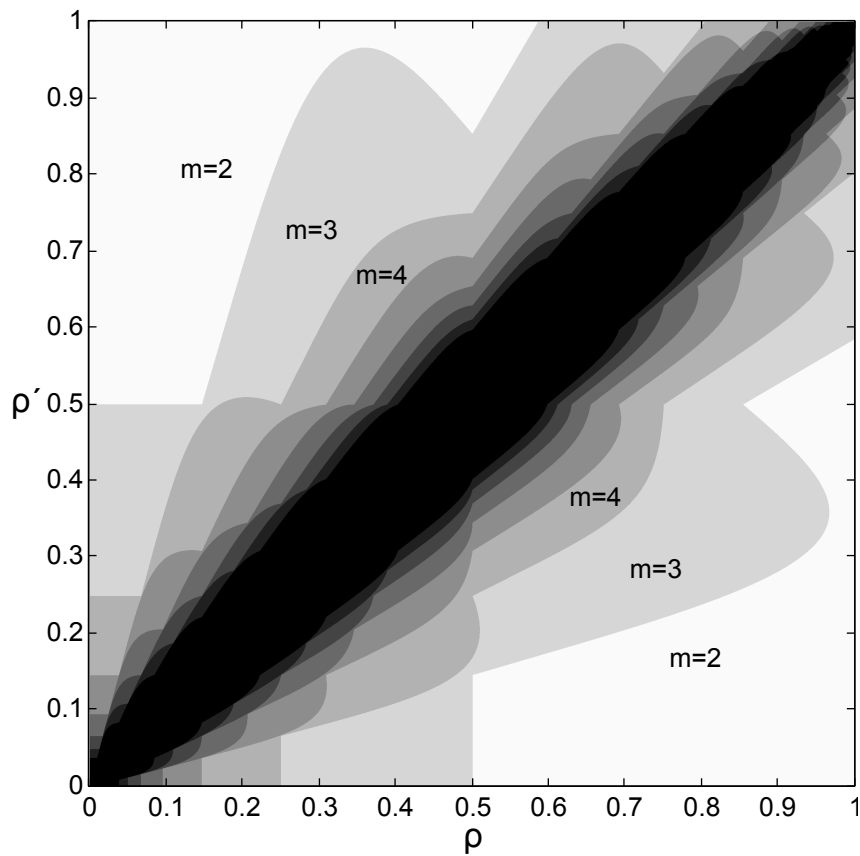}
\caption{Solutions for $m=2$ to $m=8$. Darker areas correspond to higher
$m$ values and lighter areas correspond to lower $m$ values. Lightest region is
for $m=2$. The disc-like black region on the diagonal corresponds to weight pairs
that have no solutions with $m\leq8$ iterations.}
\label{fig:p2p8}
\end{figure}

In Fig.~\ref{fig:p2p8}, numerical results showing decidability of weights
with several iterations ($m=2$ to $m=8$) are plotted. It can be seen in the
figure that, with each run, more weight combinations can be distinguished.
However the rate of addition of distinguishable weight combinations decreases
rapidly with each $m$.

Notice that for any number of iterations, the area corresponding to
distinguishable weights is symmetric with respect to $\rho = \rho^{\prime}$
line since we start with the same initial states and we take the inner
product of the final states. On the other hand we may expect one more
symmetry. The problem of distinguishing $\rho$ and $\rho^{\prime}$ should be
no different than distinguishing  $1-\rho$ and $1-\rho^{\prime}$ because
flipping function outputs does not change the complexity of the problem, thus
the figure should also have been symmetric with respect to the other
diagonal. However possible advantageous output flips does not come out
naturally as solutions in our formalism. This is also a strong indication for
the non-optimality of the solutions provided in this contribution.

A quick but rough analysis can be made for comparing the efficiencies of
classical and quantum algorithms. Classically, with a non-probabilistic
algorithm we need to make
\begin{equation}
m_{cl,det} \approx N(1-\left|\rho-\rho^\prime\right|)
\end{equation}
queries in the worst case. However we can do better with a probabilistic
algorithm. In such an algorithm, the weight can be estimated by computing the
function $f$ for $s$ random inputs. To be able to distinguish two weights
$\rho$ and $\rho^\prime$, the variance of the estimate should be smaller than
$|\rho-\rho^\prime|$. For large $N$, this process can be approximated as a
binomial process, which leads to a value of
\begin{equation}
 m_{cl,prb} \approx \frac{4\rho(1-\rho)}{\left|\rho-\rho^\prime\right|^2}
 \label{Eq:ClProbRunTime}
\end{equation}
evaluations. In the quantum case, a quick estimate of minimum evaluation
number can be obtained as follows: since the $(A,B)$ curve always starts at
the point $(1,0)$, we can only look if this curve intersects with the
negative $A$ axis. For large $m$, this approximation is good enough for
estimating the order of the run-time of the algorithm. Thus we look for the
roots of $B$ and see if the value of $A$ can be negative at these roots. As
can be seen from Eq.~\eqref{eq:ABexprA} either $m\theta_{if}$ or
$m\theta_{ig}$ is an integer multiple of $\pi$ at the roots of $B$. The value
of $A$ can be negative for these cases only if
$\left|m\theta_{if}-m\theta_{ig}\right| > \frac{\pi}{2}$. Since we are
dealing with weights that are close to each other, we can linearize the
expression and finally obtain
\begin{equation}
m_{quant} \approx \frac{2\sqrt{\rho(1-\rho)}}{\left|\rho-\rho^\prime\right|}\enskip.\label{Eq:QuantRunTime}
\end{equation}
A comparison of Equations \eqref{Eq:ClProbRunTime} and
\eqref{Eq:QuantRunTime} indicate that a quadratic speedup is obtained by
using Grover iterations for the weight distinuishability problem. This result
is compatible with the bounds given for the Grover algorithm
\cite{Bennett97,Boyer98,Zalka99,Brassard98} and Choi and Braunstein's
algorithms\cite{Braunstein07,Choi11}.

A comparison of the approaches given in \cite{Braunstein07,Choi11} and the
approach in this article is also in order. Even though both are sure-success
and achieve a square-root speedup, the algorithm given in Refs.
\onlinecite{Braunstein07} and \onlinecite{Choi11} needs
$m^{\prime}\approx(\pi/2)/\vert\rho-\rho\prime\vert$ iterations in the limit
where $\rho$ approaches $\rho^\prime$. For moderate values of the weights,
that number is only slightly bigger than this article's result of
Eq.~\eqref{Eq:QuantRunTime}. However, their algorithm specifies a complete
algorithm that solves the problem exactly, while the study presented in this
letter lacks such a clearly constructed algorithm. A special case where our
approach becomes useful is those weights where only $m=1$ or $m=2$ function
evaluations are necessary for distinguishability. On the other hand in Refs.
\onlinecite{Braunstein07} and \onlinecite{Choi11}, at least three evaluations
are needed.

\section{Conclusions\label{sec:Conclusions}}

The sure-success weight decision problem of Boolean functions using
generalized Grover iterations is discussed. Specifically, the pairs of
weights that can be distinguished by $m=1$ or $m=2$ function evaluations are
analyzed in detail. The decidability problem is reduced to a problem of
determining if a point lies in the convex hull of a curve, where the curve is
defined by polynomial functions whose order increases with the increasing
number of iterations $m$. As a result, only for cases with very small $m$ one
can obtain analytical expressions. For $m\geq3$, it becomes necessary to
follow the numerical approach.

This analysis may be compared with Braunstein and Choi's works
\cite{Braunstein07,Choi11} and with quantum counting \cite{Brassard98}.
Braunstein and Choi have shown that their algorithm is 4 times faster than
quantum counting\cite{Choi11}. Being in parallel with their result and the
known bounds in the literature, we also show that quadratic speedup is
obtained for large values of the number of iterations.

\begin{acknowledgments}
K.U. acknowledges the support of the Scientific and Technical Research
Council of Turkey (T\"{U}B\.{I}TAK). S.T. also acknowledges the support of
T\"{U}B\.{I}TAK through project 110T335.
\end{acknowledgments}


\begin{thebibliography}{10}

\bibitem{Deutsch85}D. Deutsch, Proc. R. Soc. London A \textbf{400}, 97
    (1985).

\bibitem{Deutsch92}D. Deutsch and R. Jozsa, Proc. R. Soc. London A
    \textbf{439}, 553 (1992).

\bibitem{Simon94}D. R. Simon, SIAM J. Comput. \textbf{26}, 1474 (1997).

\bibitem{Shor94}P.W. Shor, SIAM J. Comput. \textbf{26}, 1484 (1997).

\bibitem{Grover97}L. K. Grover, Phys. Rev. Lett., \textbf{79}, 325 (1997).

\bibitem{Grover98}L. K. Grover, Phys. Rev. Lett., \textbf{80}, 4329 (1998).

\bibitem{Bennett97}C. H. Bennett, E. Bernstein, G. Brassard, and U. Vazirani,
    SIAM J. Comput. \textbf{26}, 1510 (1997);

\bibitem{Boyer98}M. Boyer, G. Brassard, P. H{\o}yer, and A. Tapp, Fortsch.
    Phys. \textbf{46}, 493 (1998).

\bibitem{Zalka99} C. Zalka, Phys. Rev. A 60, 2746 (1999).

\bibitem{Biham99}E. Biham, O. Biham, D. Biron, M. Grassl, and D. Lidar, Phys.
    Rev. A \textbf{60}, 2742 (1999)

\bibitem{Biham00}E. Biham, O. Biham, D. Biron, M. Grassl, D. A. Lidar, and D.
    Shapira, Phys. Rev. A \textbf{63}, 012310 (2000).

\bibitem{Hoyer00} P. H{\o}yer, Phys. Rev. A \textbf{62}, 052304 (2000).

\bibitem{Long01} G. L. Long, Phys. Rev. A \textbf{64}, 022307 (2001).

\bibitem{Tulsi06}T. Tulsi, L. K. Grover, and A. Patel, Quantum Inf. Comput.
    \textbf{6}, 483 (2006).

\bibitem{Brassard98} G. Brassard, P. H{\o}yer, and A. Tapp, in Proceedings of
    the 25th International Colloquium on Automata, Languages and Programming,
    Lect. Notes In Comput. Sci. Vol 1443, \textbf{820}, (1998).

\bibitem{Filiol98} Filiol E. and Fontaine C., Proceedings of Advances in
    Cryptology--EUROCRYPT '98, International Conference on the Theory and
    Application of Cryptographic Techniques (Lecture Notes in Computer
    Science), \textbf{1403}, 475, (1998).

\bibitem{MacWilliams96} MacWilliams, F.J. and Sloane, N.J.A., The theory of
    error--correcting codes. North Holland (1996)

\bibitem{Chakrabarty96} Chakrabarty, K. and Hayes, J.P. Balance testing and
    balance--testable design of logic circuits. J. Electron. Testing
    \textbf{8}(1), 71, (1996).

\bibitem{Chakrabarty95} Chakrabarty, K. and Hayes, J.P. Cumulative balance
    testing of logic circuits. IEEE Trans. VLSI Syst. \textbf{3}(1), 72,
    (1995).
\bibitem{Braunstein07} S. L. Braunstein, B. S. Choi, S. Ghosh, and S. Maitra,
    J. Phys. A: Math. Theor. \textbf{40}, 8441 (2007).

\bibitem{Choi11} B. S. Choi and S. L. Braunstein, Quantum Inf. Process.
    \textbf{10}, 177 (2011).
\end{thebibliography}
\end{document}